\documentclass{ws-procs975x65}

\begin{document}

\title{New Perspective On Space And Time \\From Lorentz Violation\footnote{Plenary talk at First LeCosPA Symposium: Towards Ultimate
Understanding of the Universe (LeCosPA2012), National Taiwan
University, Taipei, Taiwan, February 6-9, 2012. }}

\author{BO-QIANG MA}

\address{School of Physics and State Key Laboratory of Nuclear Physics and
Technology, Peking University, Beijing 100871, China\\ Center for
High Energy Physics, Peking University, Beijing 100871, China
\\Center for History and Philosophy of Science, Peking
University, Beijing 100871, China\\
mabq@pku.edu.cn}

\begin{abstract}
I present a brief review on space and time in different periods of
physics, and then talk on the nature of space and time from physical
arguments. I discuss the ways to test such a new perspective on
space and time through searching for Lorentz violation in some
physical processes. I also make an introduce to a newly proposed
theory of Lorentz violation from basic considerations.
\end{abstract}


\bodymatter

\section{Space and Time in Physics}

Space and time have been discussed in human history over thousands
of years, and their nature still remains mysterious to human beings.
There are many issues concerning the nature of space and time, such
as
\begin{itemize}
\item
Whether space and time are existences or concepts?
\item
Whether they are objective or subjective?
\item
Whether they are continuous or discrete?
\end{itemize}
There are many speculations on these questions from different
perspectives, such as from metaphysics, art, philosophy, or science.
The speculations are also different in different periods of human
history.

In physics, space and time are where all physical events take place.
However, there are different understandings concerning the
properties of space and time in different periods of physics.

In classical mechanics, space is 3-dimensional and time is
1-dimensional, and they have the following properties
\begin{itemize}
\item
Space and time are independent from each other;
\item
Space and time are objective and continuous;
\item
Time is universal and observer independent;
\item
Space are observer dependent.
\end{itemize}
The space and time provide the 3+1=4 dimensions of degrees of
freedom where the Newton's laws of motion and gravity act on objects
existing in space and time. The space between different inertial
frames of reference are connected by the Galileo transformation.

In 1905, Einstein established his theory of special relativity. The
special relativity offers a revolution to the concepts of space and
time in Newton's mechanics, and provides a theoretical derivation of
adopting the Lorentz transformation for the covariance of the
equations of electrodynamics, to replace the traditional Galileo
transformation in classical physics. Space and time are unified into
4-dimensional space-time. The space and time in special relativity
have the following properties
\begin{itemize}
\item
Space and time are dependent with each other;
\item
Both space and time are observer dependent;
\item
Space and time are continuous and flat.
\end{itemize}
There are two basic principles of special relativity:
\begin{itemize}
\item
 Principle of
Relativity: the equations describing the laws of physics have the
same form in all inertial frames of reference.
\item
Principle of constant light speed: the speed of light is the same in
all directions in vacuum in all reference frames, regardless whether
the source of the light is moving or not.
\end{itemize}
These two principles lead to the unification of space and time into
a 4-dimensional space-time satisfying the Lorentz symmetry.

To unify the relativity with Newton's law of gravity, Einstein
developed his theory of general relativity during 1907-1915. Then
the curvature of space-time is determined by energy and momentum
distribution. The space and time in general relativity have the
following properties
\begin{itemize}
\item
Space and time are dependent with each other;
\item
Both space and time are observer and also matter-distribution
dependent;
\item
Space and time are continuous and can be curved.
\end{itemize}
One of the essence of general relativity is the principle of
equivalence between gravity and inertial force, and this means that
every observer can find a local inertial frame which is free from
any gravitational effect. Thus the Lorentz symmetry always holds in
such kind of local inertial frames.

Einstein's theories of relativity have been proved to be valid at
very high precision and thus have achieved great triumphs. The
Lorentz invariance, i.e., that statement that physical laws keep
invariant under the Lorentz transformation, becomes a basic
theoretical foundation of physics. Then we need to face the
question:
\begin{itemize}
\item
{Is there any reason that we seek for Lorentz violation?}
\end{itemize}

The Lorentz symmetry is a symmetry related with space and time,
therefore the Lorentz violation should be related to the basic
understandings of space and time. From the viewpoint of physics, the
origin for the breaking down of conventional concepts of space and
time might be traced back to Planck. With three fundamental
constants in physics: the Newton gravitational constant $G$, the
light speed in vacuum $c$, and the Boltzmann constant $k_B$, Planck
introduced a new constant $\hbar$ In 1899, for the purpose to
construct a ``God-given'' unit system\cite{p99}. By setting the
above four constants as bases, one can construct the Planck unit
system with a number of basic quantities, such as the Planck length
$l_{\rm P} \equiv \sqrt{G\hbar/c^3} \simeq 1.6 \times 10^{-35}$~m,
the Planck time $t_{\rm P} \equiv \sqrt{G\hbar/c^5} \simeq 5.4
\times 10^{-44}$~s, the Planck energy $E_{\rm P} \equiv \sqrt{\hbar
c^5/G} \simeq 2.0 \times 10^{9}$~J, and the Planck temperature
$T_{\rm P} \equiv \sqrt{\hbar c^5/Gk_B^2} \simeq 1.4 \times
10^{32}$~K. Therefore one may suspect that conventional
understanding of space and time might be breaking down at the Planck
scale\cite{lv5}: i.e., at the Planck length $l_{\rm P}$, or the
Planck time $t_{\rm P}$, or the Planck energy $E_{\rm P}$, where new
features of existence may emerge. The breaking down of continue
space-time was also conjectured\cite{Snyder,Wheeler}. The
expectation for the existence of a minimal length as the Planck
length led also to the establishment of some theories, e.g., the
doubly special relativity
(DSR)\cite{Amelino-Camelia2002,Magueijo:2001cr}.

Just recently, Xu and I provided a physical argument for the
discreteness of space and time\cite{xu-l}. From two known entropy
constraints:
\begin{equation}
S_{\mathrm{matter}} \leq 2\pi E R, ~~~~  \mathrm{and}
~~S_{\mathrm{matter}} \leq \frac{A}{4},
\end{equation}
combined with the black-body entropy,
\begin{equation} S =
\frac{4}{45}{\pi}^2 {T}^3 V = \left(\frac{16}{135}\right){\pi}^3 R^3
T^3,
\end{equation}
we arrive at a minimum value of space
\begin{equation}
 R\geq \Big( \frac{128}{3645\pi}\Big)^{\frac{1}{2}} l_{\mathrm{P}} \simeq 0.1
 l_{\mathrm{P}}.
 \end{equation}
Thus we reveal from physical arguments that space-time is discrete
rather than continuous. From another point of view, the newly
proposed entropic gravity suggests gravity as an emergent force
rather than a fundamental one\cite{Verlinde,HeMa}. If gravity is
emergent, a new fundamental constant should be introduced to replace
the Newtonian constant $G$\cite{lv5}. It is natural to suggest a
fundamental length scale, and such constant can be explained as the
smallest length scale of quantum space-time. Its value can be
measured through searches of Lorentz violation\cite{lv5,xu-l}. The
existence of an ``{\ae}ther" (or a ``vacuum" at rest in a specific
frame) can also bring the breaking down of Lorentz
invariance\cite{Dirac,Bjorken}.

Therefore the research on the Lorentz violation may provide us the
chance for new understanding of the nature of basic concepts, such
as ``space", ``time", and ``vacuum", through physical ways, rather
than from the perspectives of metaphysics or philosophy. It is thus
necessary to push forward the studies on Lorentz violation from both
theoretical and experimental aspects.

\section{A Glimpse on Lorentz Violation Studies}

Nowadays, there has been an increasing interest in Lorentz
invariance Violation (LV or LIV) both theoretically and
experimentally. The possible Lorentz symmetry violation effects have
been sought for from various theories, motivated by the unknown
underlying theory of quantum gravity together with various
phenomenological
applications\cite{ShaoMa10,lv3,Shao2010,Shao2011,lv4,Bietenholz:2008ni}.
This can happen in many alternative theories, e.g., the doubly
special relativity
(DSR)\cite{Amelino-Camelia2002,Magueijo:2001cr,Zhang2011}, torsion
in general relativity\cite{LV-GR1,Ni:2009fg,LV-GR2}, and large
extra-dimensions\cite{Ammosov2000,Pas2005} {\it et al.}. As
examples, I list below some phenomenological consequences of the
Lorentz violation effects studied by my students and I in the last
few years:
\begin{itemize}
\item
The Lorentz violation could provide an explanation of neutrino
oscillation without neutrino mass~\cite{Xiao08,Yang09}. We carried
out Lorentz violation contribution to neutrino oscillation by the
effective field theory for Lorentz violation and give out the
equations of neutrino oscillation probabilities. In our model,
neutrino oscillations do not have drastic oscillation at low energy
and oscillations still exist at high energy. It is possible that
neutrinos may have small mass and both Lorentz violation and the
conventional oscillation mechanisms contribute to neutrino
oscillation.

\item
The modified dispersion relation of the proton could increase the
threshold energy of photo-induced meson production of the proton and
cause an increase of the GZK cutoff energy. The earlier reports on
super-GZK events triggered attention on Lorentz-Violation. The new
results of observation of GZK cut-off put strong constraints on
Lorentz violation parameters\cite{Xiao08}.
\item
The modified dispersion relation of the photon may cause time lag of
photons with different energies when they propagate in space from
far-away astro-objects. The Lorentz violation can modify the photon
dispersion relation, and consequently the speed of light becomes
energy-dependent\cite{lv3}. This results in a tiny time delay
between high energy photons and low energy ones. Very high energy
photon emissions from cosmological distance can amplify these tiny
LV effects into observable quantities. We analyzed photons from
$\gamma$-ray bursts from Fermi satellite observations and presented
a first robust analysis of these taking the intrinsic time lag
caused by sources into account, and gave an estimate to LV energy
scale $\sim 2 \times 10^{17}$~GeV for linear energy dependence, and
$\sim 5 \times 10^9$~GeV for quadratic dependence\cite{Shao2010}.
\item
We also studied recent data on Lorentz violation induced vacuum
birefringence from astrophysical consequences\cite{Shao2011}. Due to
the Lorentz violation, two helicities of a photon have different
phase velocities and group velocities, termed as ``vacuum
birefringence''. From recently observed $\gamma$-ray polarization
from Cygnus X-1, we obtained an upper limit $\sim 8.7\times10^{-12}$
for Lorentz-violating parameter $\chi$, which is the most firm
constraint from well-known systems.
\end{itemize}

\section{A Newly Proposed Theory of Lorentz Violation from Basic Principles}

Among many theoretical investigations of Lorentz violation, it is a
powerful framework to discuss various LV effects based on
traditional techniques of effective field theory in particle
physics. Here we focus our attention on a newly proposed theory of
Lorentz violation from basic principles: the Standard Model
Supplement (SMS)\cite{Ma10,SMS3}.

It is clear that human should not be narrowly focused just on
effective theories with some additional terms beyond the
conventional theory added by hand. It is a basic requirement that we
should find a fundamental theory to derive the Lorentz violation
terms from basic consideration. In the standard model supplement
(SMS) framework\cite{Ma10,SMS3}, the LV terms are brought about from
a basic principle denoted as the physical independence or physical
invariance (PI):
\begin{itemize}
\item
Principle of Physical Invariance: the equations describing the laws
of physics have the same form in {\bf all admissible mathematical
manifolds}.
\end{itemize}
The principle leads to the following replacement of the ordinary
partial $\partial_{\alpha}$ and the covariant derivative
$D_{\alpha}$
\begin{equation}\label{eqn:substitution}
\partial^{\alpha} \rightarrow M^{\alpha\beta}\partial_{\beta},\quad
D^{\alpha}\rightarrow M^{\alpha\beta}D_{\beta},
\end{equation}
where $M^{\alpha\beta}$ is a local matrix. The Lorentz violation
terms are thus uniquely determined from the standard model
Lagrangian without any ambiguity\cite{Ma10}, and their general
existence is derived from basic consideration rather than added by
hand. The explicit form of the matrices $M^{\alpha\beta}$ demands
more basic theories concerning the true nature of space and time,
and we suggest to adopt a physical way to explore these matrices
through experiments rather than from theory at first. For more
generality, we do not make any ad hoc assumption about these
matrices. Thus these matrices might be particle dependent
corresponding to the standard model particles under consideration,
with the elements of these matrices to be measured or constrained
from experimental observations.

We separate $M^{\alpha\beta}$ to two matrices like $M^{\alpha
\beta}=g^{\alpha \beta}+\Delta^{\alpha \beta}$, where
$g^{\alpha\beta}$ is the metric tensor of space-time and
$\Delta^{\alpha \beta}$ is a new matrix which is particle-type
dependent generally. Since $g^{\alpha\beta}$ is Lorentz invariant,
$\Delta^{\alpha \beta}$ contains all the Lorentz violating degrees
of freedom from $M^{\alpha \beta}$. Then $\Delta^{\alpha \beta}$
brings new terms violating Lorentz invariance in the standard model
and is called Lorentz violation matrix. The theory returns back to
the standard model when these Lorentz violation matrices vanish.

More explicitly, the effective Lagrangian
$\mathcal{L}_{\mathrm{SM}}$ of the minimal standard model is
composed of four parts
\begin{eqnarray}
\mathcal{L}_{\mathrm{SM}}&=& \mathcal{L}_{\mathrm{G}}+\mathcal{L}_{\mathrm{F}}+\mathcal{L}_{\mathrm{H}}+\mathcal{L}_{\mathrm{HF}},\label{eqn:SM}\\
\mathcal{L}_{\mathrm{G}} &=& -\frac{1}{4}F^{a\alpha\beta } F_{\alpha\beta}^{a}, \label{eqn:SMG}\\
\mathcal{L}_{\mathrm{F}} &=&i\bar{\psi}\gamma^{\alpha}D_{\alpha}\psi, \label{eqn:SMF}\\
\mathcal{L}_{\mathrm{H}}&=& (D^{\alpha}\phi)^{\dag}D_{\alpha}\phi +
V(\phi),\label{eqn:SMHG}
\end{eqnarray}
where 
we omit the chiral differences, the summation of chirality and gauge
scripts.
 $\psi$ is the fermion field, $\phi$ is the Higgs
field, and $V(\phi)$ is the Higgs self-interaction.
$F_{\alpha\beta}^{a}=\partial_{\alpha}A_{\beta}^{a}-\partial_{\beta}A_{\alpha}^{a}
-gf^{abc}A_{\alpha}^{b}A_{\beta}^{c}$,
$D_{\alpha}=\partial_{\alpha}+igA_{\alpha}$ and
$A_{\alpha}=A_{\alpha}^{a}t^{a}$, with $A_{\alpha}^{a}$ being the
gauge field. $g$ is the coupling constant, and $f^{abc}$ and $t^{a}$
are the structure constants and generators of the corresponding
gauge group respectively. $\mathcal{L}_{\mathrm{HF}}$ is the Yukawa
coupling between the fermions and the Higgs field, and is not
related to derivatives, thus it remains unchanged under the
replacement (\ref{eqn:substitution}).

Under (\ref{eqn:substitution}) and the decomposition
$M^{\alpha\beta}=g^{\alpha\beta}+\Delta^{\alpha\beta}$, the
Lagrangians in (\ref{eqn:SMG})-(\ref{eqn:SMHG}) become
\begin{eqnarray}
\mathcal{L}_{\mathrm{G}} &=& -\frac{1}{4}
 (M^{\alpha\mu}\partial_{\mu}A^{a\beta}-M^{\beta\mu}\partial_{\mu}A^{a\alpha}
  -gf^{abc}A^{b\alpha}A^{c\beta})       \nonumber\\
  &\times&
 (M_{\alpha\mu}\partial^{\mu}A_{\beta}^{a}-M_{\beta\mu}\partial^{\mu}A_{\alpha}^{a}
  -gf^{abc}A_{\alpha}^{b}A_{\beta}^{c}) \nonumber\\
  &=&-\frac{1}{4}F^{a\alpha\beta }F_{\alpha\beta}^{a}+ \mathcal{L}_{\mathrm{GV}},\label{eqn:SMSG}\\
\mathcal{L}_{\mathrm{F}}
&=&i\bar{\psi}\gamma_{\alpha}M^{\alpha\beta}D_{\beta}\psi
  =i\bar{\psi}\gamma^{\alpha}D_{\alpha}\psi+  \mathcal{L}_{\mathrm{FV}}, \label{eqn:SMSF} \\
\mathcal{L}_{\mathrm{H}}&=&(M^{\alpha\mu}D_{\mu}\phi)^{\dag}M_{\alpha\nu}D^{\nu}\phi
  + V(\phi)                                                                \nonumber \\
  &=&(D^{\alpha}\phi)^{\dag}D_{\alpha}\phi + V(\phi)+ \mathcal{L}_{\mathrm{HV}},\label{eqn:SMSHG}
\end{eqnarray}
with $M^{\alpha\beta}$ being the real matrix to maintain the
Lagrangian hermitian. The last three terms
$\mathcal{L}_{\mathrm{GV}}$, $\mathcal{L}_{\mathrm{FV}}$ and
$\mathcal{L}_{\mathrm{HV}}$ of the equations mentioned above are the
supplementary terms for the ordinary Standard Model. The explicit
forms of these terms are
\begin{eqnarray}
\mathcal{L}_{\mathrm{GV}}&=&
-\frac{1}{2}\Delta^{\alpha\beta}\Delta^{\mu\nu}(g_{\alpha\mu}\partial_{\beta}
A^{a\rho}\partial_{\nu}A_{\rho}^{a}-\partial_{\beta}A_{\mu}^{a}
\partial_{\nu}A_{\alpha}^{a})    \nonumber \\
   & & -F_{\mu\nu}^{a}\Delta^{\mu\alpha}\partial_{\alpha}A^{a\nu},\label{eqn:GV}  \\
\mathcal{L}_{\mathrm{FV}}&=&
 i\Delta^{\alpha\beta}\bar{\psi}\gamma_{\alpha}\partial_{\beta}\psi
  -g\Delta^{\alpha\beta}\bar{\psi}\gamma_{\alpha}A_{\beta}\psi, \label{eqn:FV} \\
\mathcal{L}_{\mathrm{HV}}&=&
(g_{\alpha\mu}\Delta^{\alpha\beta}\Delta^{\mu\nu}+\Delta^{\beta\nu}+\Delta^{\nu\beta})
  (D_{\beta}\phi)^{\dag}D_{\nu}\phi.  \label{eqn:HGV}
\end{eqnarray}
Thus we obtain a new effective Lagrangian for the Standard Model
with new supplementary terms, denoted by
$\mathcal{L}_{\mathrm{SMS}}$
\begin{eqnarray}
&\mathcal{L}_{\mathrm{SMS}}&= \mathcal{L}_{\mathrm{SM}} + \mathcal{L}_{\mathrm{LV}}, \label{eqn:SMS}\\
&\mathcal{L}_{\mathrm{LV}}&=
\mathcal{L}_{\mathrm{GV}}+\mathcal{L}_{\mathrm{FV}}+\mathcal{L}_{\mathrm{HV}},\label{eqn:LIV}
\end{eqnarray}
where $\mathcal{L}_{\mathrm{SMS}}$ satisfies the Lorentz covariance
(SO$^{+}$(1,3)), the gauge symmetry invariance of
SU(3)$\times$SU(2)$\times$U(1) and invariance under the requirement
of the principle of physical invariance or independence (PI), under
which $\mathcal{L}_{\mathrm{SM}}$ cannot remain unchanged in a
general situation.

We can have a better understanding of the LV terms in SMS here. The
elements of $M^{\alpha\beta}$ of a particle are mass dimensionless
(which is natural for the sign of testifying the Lorentz
invariance), and they are not global constants generally. All of the
LV terms are expressed in $\mathcal{L}_{\mathrm{LV}}$, and the LV
information is measured by the concise matrix
$\Delta^{\alpha\beta}$, which is convenient for a systematic study
of the LV effects. To determine whether the Lorentz invariance holds
exactly, further work is needed to analyze the effective Lagrangian
(\ref{eqn:SMS}) of QED, QCD and EW (ElectroWeak) fields, and more
experiments are needed to determine the magnitude of the elements in
the matrices $M^{\alpha\beta}$ for different particles. There have
been some preliminary progress along this line with the theory of
SMS applied to discuss the Lorentz violation effects for
protons~\cite{Ma10}, photons~\cite{SMS3}, and
neutrinos~\cite{SMS-OPERA}. More works are still needed for
systematic studies.

Generally, $\Delta^{\alpha\beta}$ might be particle-type and
flavor-type dependent. If we use the vacuum expectation values of a
$\Delta^{\alpha\beta}$ for the coupling constants in the
corresponding effective Lagrangian, not all of the 16 degrees of
freedom of $M^{\alpha\beta}$ are physical. For the derivative field
$M(\partial_{x})\varphi(x)$ of an arbitrary given field,
$\varphi(x)$ can be rescaled to absorb one of the 16 degrees of
freedom so that only 15 are left. When more fields are involved,
there is only one degree of freedom that can be reduced from a
rescaling consideration for all fields. Thus for generality, we may
keep all 16 degrees of freedom in $M^{\alpha\beta}$ for a specific
particle in our study.

The matrix $M^{\alpha\beta}$ in the SMS theory just appears with the
derivative terms of Lagrangians, but not with the coordinate terms.
Therefore one should not confuse this matrix with the metric of the
general relativity. To define a covariant derivative in general
relativity, one needs to introduce the concept of connection to
reflect the effect due to the curvature of space and time from
gravity. Our introduction of the matrix $M^{\alpha\beta}$ can be
considered as an alternative choice along the similar philosophy,
but with a more general sense by including also possible effects
from other interactions other than solely gravity. Therefore our
matrix $M^{\alpha\beta}$ can contain the effect due to general
relativity or more beyond that, but we do not intend to derive it
from theory but to detect it from phenomenological manifestations.
This is from the consideration that the space-time structure of
nature might be more complicated than just the effect from gravity.

There still exists the question of how to understand and handle the
Lorentz violation matrix $\Delta^{\alpha\beta}$. We list here three
options for understandings and treatments\cite{Ma:2011jj}:

\begin{itemize}
\item
{\bf Scenario I}: which can be called as fixed scenario in which the
Lorentz violation matrices are taken as constant matrices in any
inertial frame of reference the observer is working. It means that
the Lorentz violation matrices can be taken as approximately the
same for any working reference frames such as the earth-rest frame,
the sun-rest frame, or the CMB frame. However, there will be the
problem of inconsistency for an ``absolute physical event" between
different reference frames\cite{Ma:2012zd}, if one sticks to this
scenario. Therefore this scenario can be adopted as a practical
approach when one is focused on the Lorentz violation effect within
a certain frame and does not care about relationships between
different frames.


\item
{\bf Scenario II}: which can be called as ``new {\ae}ther" scenario
in which the Lorentz violation matrices transform as tensors between
different inertial frames but keep as constant matrices within the
same frame. The Lorentz violation matrices play the roles for the
exitance of some kinds of background fields, or the ``new {\ae}ther"
(i.e., a ``vacuum" at rest in a specific frame), which changes from
one frame to another frame by Lorentz transformation. Thus it can be
considered as a standard viewpoint to treat the matrix
$M^{\alpha\beta}$ as a tensor satisfying Lorentz symmetry.

\item
{\bf Scenario III}: which can be called as covariant scenario in
which the Lorentz violation matrices transform as tensors adhered
with the corresponding standard model particles. It means that these
Lorentz violation matrices are emergent and covariant with their
standard model particles. Such a scenario still needs to be checked
for consistency and for applications in future.
\end{itemize}

Before accepting the SMS as a fundamental theory, one can take the
SMS as an effective framework for phenomenological applications by
confronting with various experiments to determine and/or constrain
the Lorentz violation matrioce $\Delta^{\alpha \beta}$ for various
particles. So our idea is to reveal the real structure of Lorentz
violation of nature from experiments rather than from theory. We
consider this phenomenological way as more appropriate for physical
investigations, rather than to derive everything from theory at
first. As a comparison, the specific form of the quark mixing matrix
is determined from experimental measurements rather than derived
from theory\cite{Ma:2011gh}. Even after so many years of research
and also the elements of the CKM mixing matrix have been measured to
very high precision, there is still no a commonly accepted theory to
derive the quark mixing matrix from basic principles.

We now provide some remarks concerning the Lorentz violation studies
in field theory frameworks. In the effective field theory
frameworks, the standard particles transform according to the
Lorentz symmetry between different momentum states. The background
fields should also transform according to the Lorentz symmetry
between different observer working frames from the requirement of
consistency. From this sense, there is actually no Lorentz violation
for the whole system of the standard model particles together with
the background fields. The Lorentz violation exists for the standard
model particles within an observer working frame, when these
particles have different momenta between each other. From this
sense, the Lorentz violation is due to the existence of the
background fields, which are treated as fixed parameters in the
observer working frame.

The newly proposed theory of SMS not only provides clear
relationship between some general LV parameters\cite{SMS-photon2},
but also can be conveniently applied for phenomenological
analysis\cite{Ma10graal}. We would need more experimental
investigations to check whether it can meet the criterion of being
able to provide a satisfactory description of the physical reality,
with simplicity and beauty in formalism, together with the
predictive power towards new knowledge for human beings. It is also
possible that the nature satisfies the Lorentz symmetry perfectly
and we would be unable to find a physical evidence to support the
theory. This would imply that the newly introduced Lorentz violation
matrix $\Delta^{\alpha\beta}$ would vanish for nature.

\section{Conclusion}

From the discussions above, we may have some new perspectives on the
nature of space-time:
\begin{itemize}
\item
Space and time are 3+1=4 dimensional or may have extra-dimensions;
\item
Both space and time are observer and object dependent;
\item
Space and time might be particle-type dependent and can be curved by
the existence of matter.
\item
Space and time might be discrete and such discreteness can be tested
through experiments on Lorentz violation.
\end{itemize}

\noindent Finally we present our conclusion:
\begin{itemize}
\item
Researches on Lorentz violation have been active for many years;
\item
There might be some marginal evidences for Lorentz violation yet,
but non of them can be considered as confirmed;
\item
The Lorentz violation study can bring conceptual revolution on the
understanding of space-time for human beings;
\item
Lorentz violation is being an active frontier both theoretically and
experimentally.
\end{itemize}

\section*{Acknowledgements}
I acknowledge Professor Pisin Chen for his warm invitation and
hospitality for attending the 1st LeCosPA Symposium. I am very
grateful for the discussions and collaborations with a number of my
students: Zhi Xiao, Shi-Min Yang, Lijing Shao, Lingli Zhou, Xinyu
Zhang, Yunqi Xu, and Nan Qin, who devoted their wisdoms and
enthusiasms bravely on the topic of Lorentz violation in past few
years. The work was supported by National Natural Science Foundation
of China (Nos.~10975003, 11021092, 11035003 and 11120101004).


\end{document}